\documentclass[conference]{IEEEtran}
\usepackage{cite}
\usepackage{amsmath,amssymb,amsfonts}
\usepackage{algorithmic}
\usepackage{graphicx}
\usepackage{textcomp}
\usepackage{xcolor}
\usepackage{hyperref}
\usepackage{subcaption}

\def\BibTeX{{\rm B\kern-.05em{\sc i\kern-.025em b}\kern-.08em
    T\kern-.1667em\lower.7ex\hbox{E}\kern-.125emX}}
\begin{document}

\title{
Guiding Wireless Signals with Arrays of Metallic Linear Fresnel Reflectors: A Low-cost, Frequency-versatile, and Practical Approach
% Enhancing Wireless Signal Quality using Metallic Linear Fresnel Reflector Array
}

\author{\IEEEauthorblockN{Hieu Le\IEEEauthorrefmark{1}}
\IEEEauthorblockA{\textit{Electrical and Computer Engineering} \\
\textit{Texas A\&M University}\\
College Station, Texas, USA \\
hieult@tamu.edu}
\\
\IEEEauthorblockN{Jian Tao}
\IEEEauthorblockA{\textit{School of Performance, } \\
\textit{Visualization, and Fine Arts} \\
\textit{Texas A\&M University}\\
College Station, Texas, USA \\
jtao@tamu.edu }
\and
\IEEEauthorblockN{Oguz Bedir\IEEEauthorrefmark{1}}
\IEEEauthorblockA{\textit{Electrical and Computer Engineering} \\
\textit{Texas A\&M University}\\
College Station, Texas, USA \\
oguzbedir@tamu.edu }
\\
\IEEEauthorblockN{Sabit Ekin}
\IEEEauthorblockA{\textit{Engineering Technology, and} \\
\textit{Electrical and Computer Engineering} \\
\textit{Texas A\&M University}\\
College Station, Texas, USA \\
sabitekin@tamu.edu }
\and
\IEEEauthorblockN{Mostafa Ibrahim}
\IEEEauthorblockA{\textit{Engineering Technology } \\
\textit{and Industrial Distribution} \\
\textit{Texas A\&M University}\\
College Station, Texas, USA \\
mostafa.ibrahim@tamu.edu }
% \and
% \and
}

\maketitle
\begingroup\renewcommand\thefootnote{\IEEEauthorrefmark{1}}
\footnotetext{These authors contributed equally to this work.}
\endgroup

\begin{abstract}

This study presents a novel mechanical metallic reflector array to guide wireless signals to the point of interest, thereby enhancing received signal quality. Comprised of numerous individual units, this device, which acts as a linear Fresnel reflector (LFR), facilitates the reflection of incoming signals to a desired location. Leveraging geometric principles, we present a systematic approach for redirecting beams from an Access Point (AP) toward User Equipment (UE) positions. This methodology is geared towards optimizing beam allocation, thereby maximizing the number of beams directed towards the UE. Ray tracing simulations conducted for two 3D wireless communication scenarios demonstrate significant increases in path gains and received signal strengths (RSS) by at least 50 dB with strategically positioned devices.
% Comparative analyses between scenarios with and without devices illustrate significant improvements in path gain and RSS.

\end{abstract}

\begin{IEEEkeywords}
reconfigurable intelligent surfaces (RIS), specular reflections, path gain, received signal strength (RSS), ray tracing, coverage map
\end{IEEEkeywords}

\section{Introduction}

Addressing the challenges posed by the propagation channel has been a crucial element in achieving reliable, high-quality, and high-capacity wireless communication. Consequently, throughout history, research efforts in wireless communication have primarily focused on controlling the interaction of system endpoints—the transmitter and the receiver—with an emphasis on designing systems capable of adapting to and mitigating challenges posed by the channel \cite{bjornson2022reconfigurable, direnzo:2020}. However, focusing solely on controlling how system endpoints interact has its limitations in fully meeting the ambitious goals set for both current and future wireless networks. Thus, in conjunction with the progress in metamaterials, the innovative concept of smart radio environments \cite{direnzo:2020} has emerged—environments that can adapt based on communication requirements. This notion has garnered significant interest from the wireless research community owing to its substantial potential.

Prior to the widespread adoption of the current notion of smart radio environments, early studies explored the use of passive frequency-selective surfaces (FSS) to manipulate the environment and, in turn, control the propagation characteristics \cite{sung:2006, raspopoulos:2011}. Passive FSS, functioning as EM metamaterials, exhibit specific characteristics, e.g., reflection or transmission, depending on the frequency of the impinging EM wave. However, their limitation lies in dynamic channel characteristics, prompting the exploration of active FSS \cite{anwar:2018}. An early study \cite{subrt:2010} proposed the deployment of active FSSs on walls to dynamically adjust transmission and reflection characteristics based on communication needs. The concept envisions intelligent walls as part of a self-configuring autonomous infrastructure for dynamic FSS control. The active FSS employs positive-intrinsic-negative (PIN) diodes, allowing controlled energy passage or reflection based on the diode state. The study illustrates, through simulations, that intelligent walls can effectively control signal coverage and interference.

Another pivotal contribution \cite{liaskos:2018} presents a parallel concept, suggesting the application of software-controlled metasurfaces, termed HyperSurfaces \cite{tsioliaridou:2017}, on objects. These metasurfaces utilize meta-atoms with switches to execute absorption, steering, and polarization of impinging EM waves.

Currently, the primary research focus revolves around the enabling technology called the reconfigurable intelligent surface (RIS) for realizing smart radio environments \cite{bjornson2022reconfigurable, direnzo:2020}. These devices are constructed using electrically thin unit cells also referred to as scatterers, meta-atoms, reflecting elements in the literature, conceptualized as two-dimensional (2D) structures, i.e., surfaces. Voltage-controlled reconfiguration can be realized using a variety of tunable components, such as diodes, transistors, micro-electromechanical systems (MEMS), graphene, and liquid crystals \cite{zahra:2021}. In common implementations, however, each unit cell is equipped with either a PIN diode or a varactor diode, preferred for their swift response times, minimal reflection loss, cost-effective hardware, and energy-efficient operation \cite{basharat:2022}. Manipulation of the unit cell's impedance is achieved by adjusting the bias voltage of these diodes, involving binary control (on/off) for PIN diodes and continuous control for varactor diodes, although in practice, the control may be discrete with several levels due to imperfections and sensitivity to input bias. This modulation in impedance allows for control over the amplitude and phase of the incident EM wave, enabling functionalities like anomalous reflections and beamforming (for a more detailed exploration of EM-based elementary functions of RIS, please refer to \cite{direnzo:2020}). This approach to reconfiguring metasurfaces is just one method among several for influencing incident EM waves. Other approaches involve optical, thermal, and mechanical means, as well as power control using non-linear metasurface designs \cite{zahra:2021}.

For next-generation communications like 5G and beyond, utilizing millimeter (mmWave) and terahertz (THz) frequency bands is considered to alleviate frequency congestion. However, these high-frequency bands suffer from significant path loss and attenuation, particularly in non-line-of-sight (NLOS) scenarios. Among various emerging technologies, especially RIS has gained attention for enhancing NLOS communication performance in mmWave \& THz bands. Despite their promising capabilities, current RIS implementations are complicated and costly, and they operate within specific frequency bands due to their diffuse reflection properties \cite{jeong:2022}. This necessitates the use of different RIS designs for different frequency bands, e.g., 28 GHz, 40 GHz, and 140 GHz bands, adding to the deployment challenges. 

Drawing inspiration from \cite{khawaja:2020} and linear Fresnel reflector (LFR) concept, we propose the use of metallic surfaces to enhance the coverage and received signal strength (RSS). Unlike fixed structures, our approach allows each metallic surface that composes the reflecting device to adjust its polar angle in the $x-y$ plane and the angle above (and below) the $x-y$ plane, as illustrated in Figure \ref{figure:freshno}, \ref{figure:reflector}, and \ref{figure:specular_reflection}. This capability enables the configuration of azimuth and elevation angles to enhance the path gain based on the positions of the (access point) AP and the (user equipment) UE. 

Unlike previously mentioned RIS reconfiguration techniques, our metal-based reflector arrays offer versatility across a broad spectrum of signal frequencies. This characteristic potentially enables cost-effective deployment, as our arrays can be utilized across various wireless frequency bands. Moreover, different from traditional voltage-controlled RIS implementations that need a constant bias voltage to maintain the operation of their components, our method does not consume power once configured. 
% Consequently, our proposed approach is potentially suitable for applications where receiving devices are typically stionary, such as indoor environments and back haul extension for rural areas without the need of connecting cable, as our reflectors remain functional without the need for further adjustments after the initial setup.
Consequently, our proposed approach is well-suited for applications where receiving devices are typically stationary, such as indoor environments or backhaul extension for rural areas without the need for connecting cables.

The rest of this paper is organized as follows: Section~\ref{sec:system_model} provides a comprehensive overview of the system model, assumptions, the simulation scenarios, and their setup, and the mathematical formulation that is necessary to configure the tiles of the reflector array. Section~\ref{sec:results_and_discussion} presents the results obtained from the simulation analysis. Lastly, Section~\ref{sec:conclusion} concludes the paper with a brief summary and future research directions.

\section{System Model}
\label{sec:system_model}

\begin{figure}[h]
  \centering
  \includegraphics[width=1\linewidth]{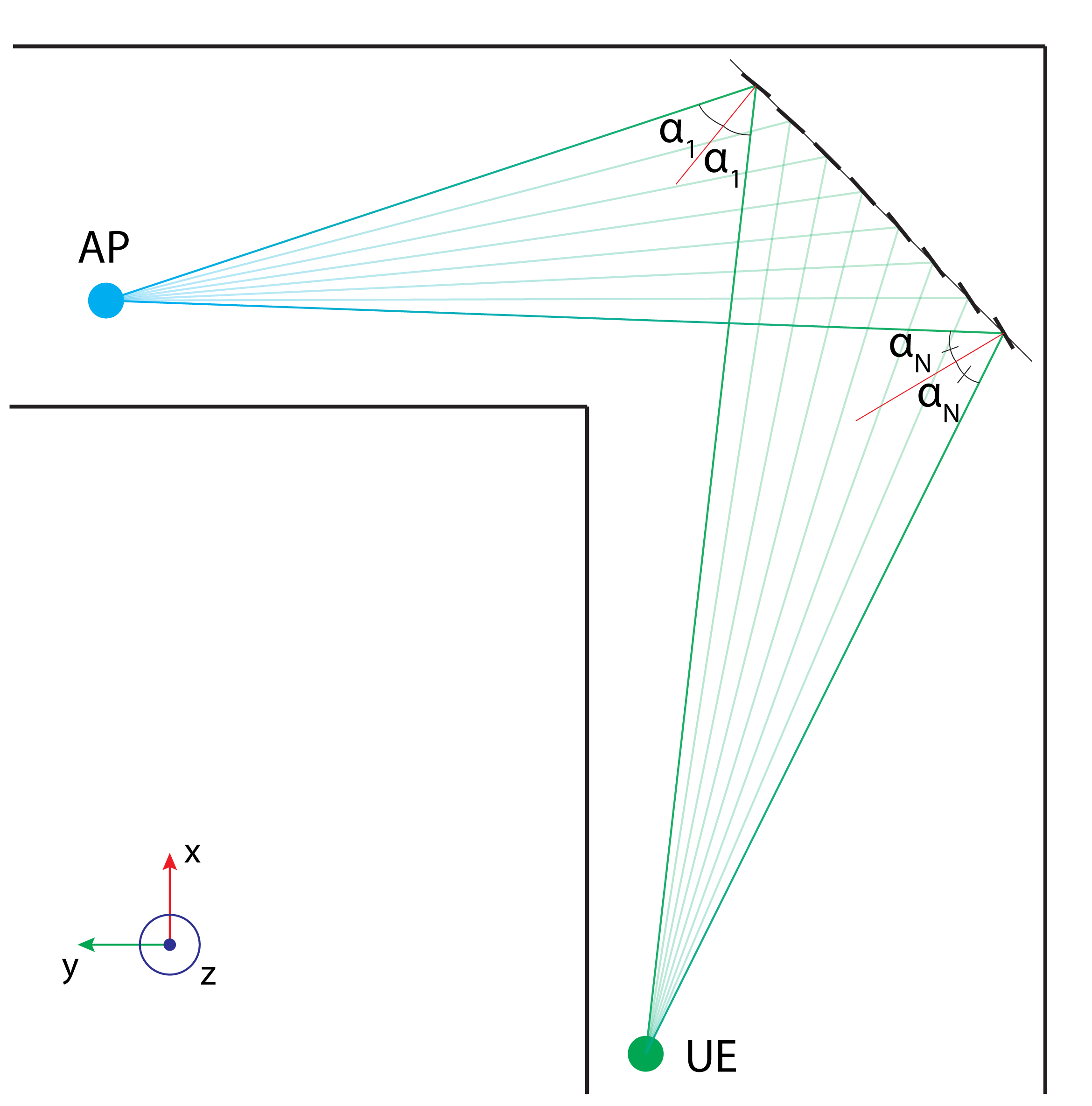}
  \caption{Metallic linear Fresnel reflector (LFR) arrays capable of steering signal beams from an AP to a UE with the bird-eye view (2D plane x-y).
  }
  \label{figure:freshno}
\end{figure}

We introduce a metal-based LFR array engineered to guide signals toward specific destinations through a beam-focusing method illustrated in Figure \ref{figure:freshno}, which relies on geometric principles. We examine various scenarios to evaluate the effectiveness of our technique. Furthermore, we elaborate on the simulation method utilized in our study.

\subsection{Metallic Linear Fresnel Reflector Array}

\begin{figure}[h]
  \centering
  \includegraphics[width=0.65\linewidth]{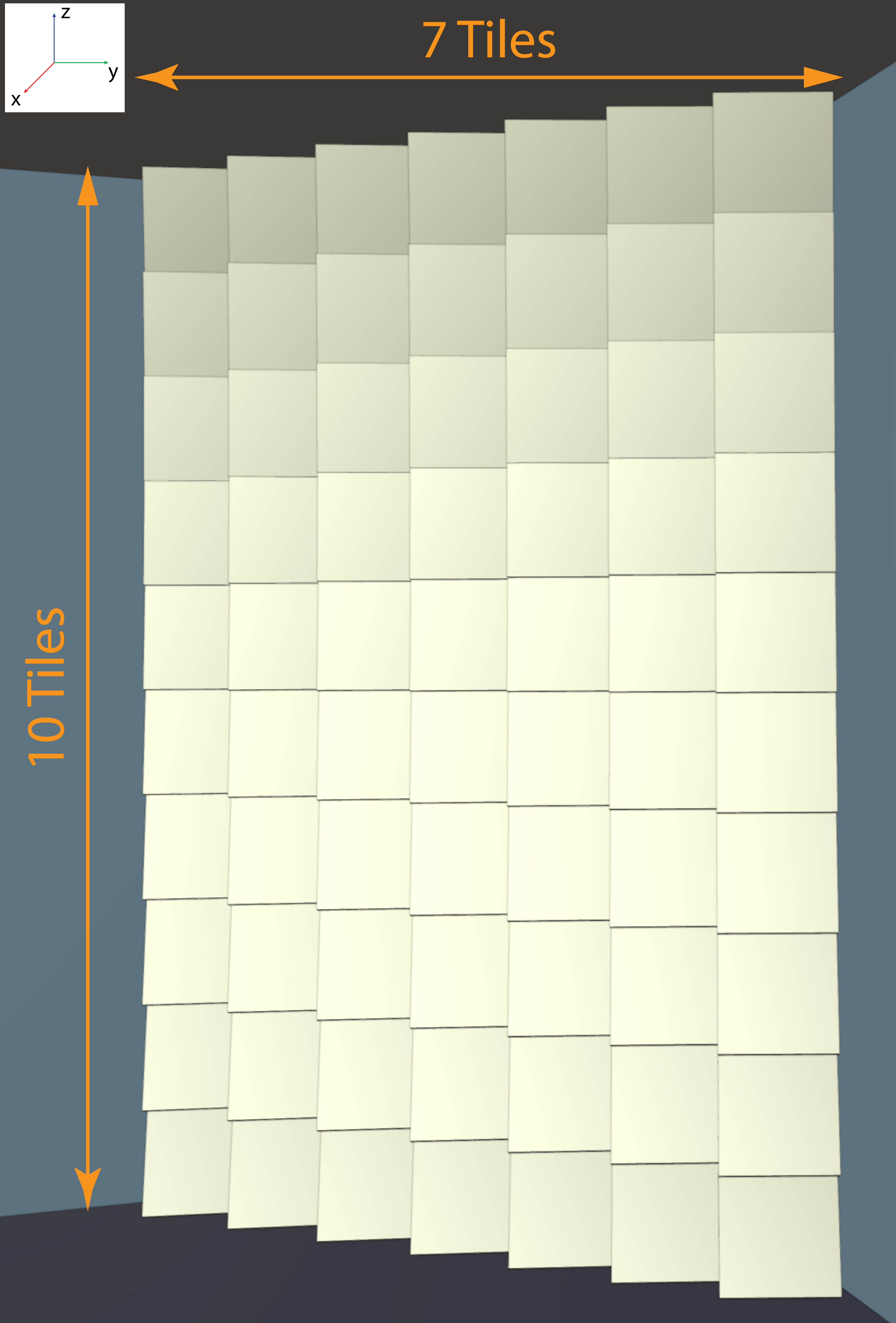}
  \caption{A metallic linear Fresnel reflector array which consists of 10 rows and 7 columns of small metallic units, called tiles.}
  \label{figure:reflector}
\end{figure}

We employ multiple specialized metallic LFR arrays designed to guide signal wavefronts back into the environment and toward a desired location. As can be seen from Figure \ref{figure:reflector}, each of our reflector arrays comprises many individual units, referred to as tiles. These tiles have the ability to rotate independently, allowing for the manipulation of the signal wavefront reflection.

\begin{figure}[h]
  \centering
  \includegraphics[width=1\linewidth]{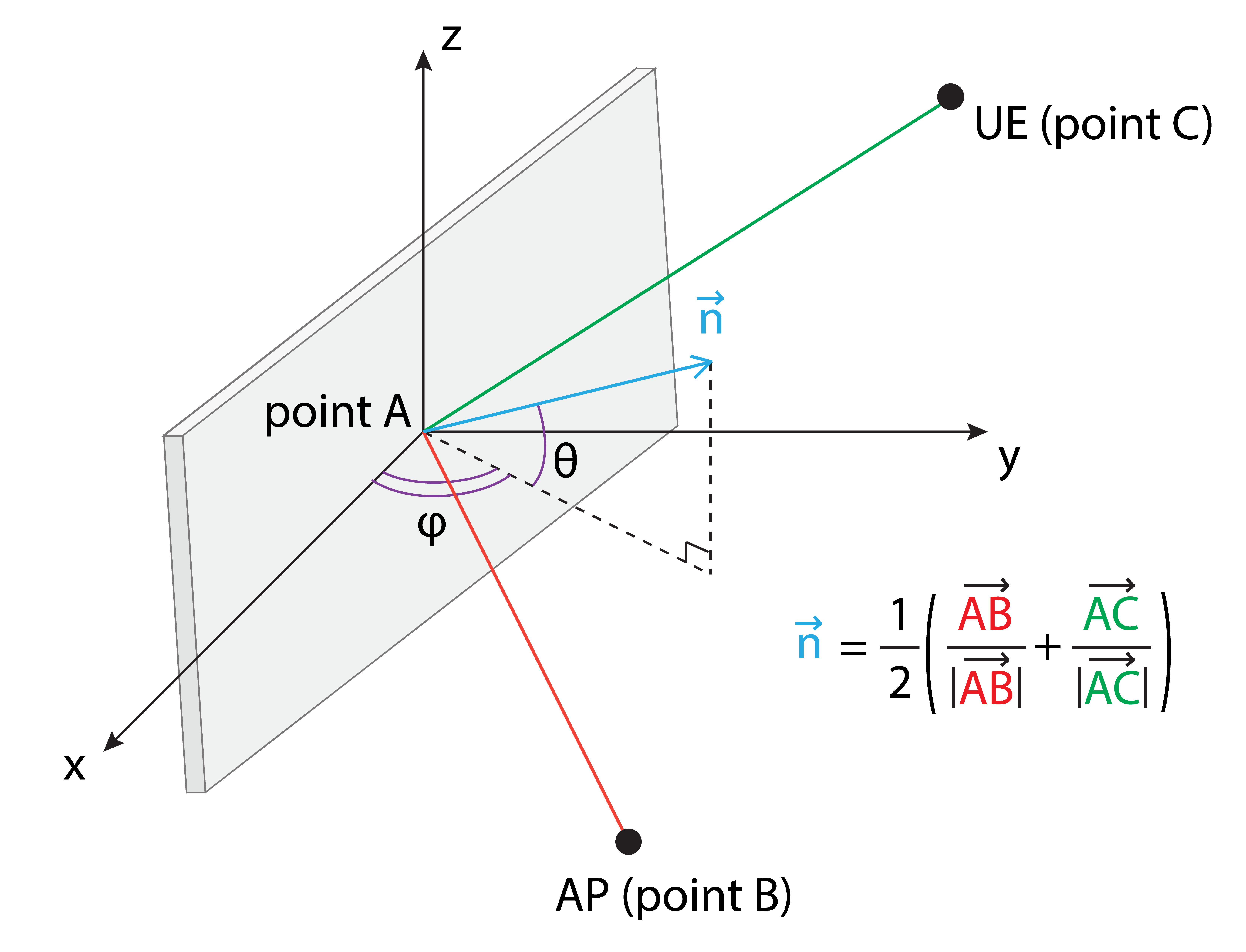}
  \captionof{figure}{A ray originating from an Access Point (Point B) intersects a surface at Point A and undergoes specular reflection towards a User Equipment located at Point C. $\vec{n}$ is the normal of the material's surface.}
  \label{figure:specular_reflection}
\end{figure}

\subsection{Beamfocusing}
\label{subsection:beamfocusing}

Our objective is to optimize signal strength at the UE position. This involves maximizing the number of transmission paths from the AP to the UE through the adjustment of metallic LFR arrays. Our approach focuses on aligning each reflector's tile to create a path from the AP to the central point of the tile and subsequently to the UE location. This strategy guarantees the existence of at least one reflected path from the AP to the UE for every tile on the reflector array. Additionally, as the number of tiles on the reflector increases, so does the quantity of reflected paths available.

% In the absence of rotation angle constraints, we have the flexibility to select rotation angles for reflector's tiles without limitations. Utilizing Snell's law, which governs that the incident angle equals the reflected angle, we aim to adjust the normal of the material's surface to achieve the equal angles.
Without constraints on rotation angles, we have the flexibility to select rotation angles for the reflector's tiles without limitations. By applying Snell's law, which governs that the angle of incidence equals the angle of reflection, our objective is to manipulate the normal of the material's surface to achieve these equal angles.
% we aim to align the normal of the material's surface as a bisector of the triangle formed by the points AP-Reflector-UE. To accomplish this objective, we approach the problem by initially assuming the normal serves as the bisector and then proceed to determine the appropriate rotation angle for each reflector element. 
For clarity, we introduce some annotations: point A represents the center of a reflector tile, point B denotes the location of the AP, point C is the UE location, and $\vec{n}$ is the normal of the material's surface (Figure \ref{figure:specular_reflection}).

We can find the normal $\vec{n}$ of the surface that satisfies Snell's law by using (\ref{eq:normal}).

\begin{equation}
    \vec{n} = \frac{1}{2} \left(\frac{\overrightarrow{AB}}{|AB|} + \frac{\overrightarrow{AC}}{|AC|}\right)
    \label{eq:normal}
\end{equation}

Initially, the tile is positioned flat on the xy-plane with its normal directed upwards along the positive z-axis. 
% Our objective now is to orient the normal towards point D. This requires modifications to the elevation angle $\theta$ and azimuth angle $\phi$. 
% While transforming the entire coordinate system to relocate point D to the origin is required to find these two angles, we recognize that only the coordinates of point A, the tile's center, need to be adjusted. Consequently, we simplify the process by solely applying a transformation to point A, which involves determining the vector $\overrightarrow{DA}$.
We use the standard transformation from the Cartesian coordinate system to the spherical coordinate system to determine both angles, $\theta$ and $\phi$ using (\ref{eq:cartesian_to_spherical_r}), (\ref{eq:cartesian_to_spherical_theta}), and (\ref{eq:cartesian_to_spherical_phi}).

\begin{equation}
    r = \sqrt{x^2 + y^2 + z^2} \text{,}
    \label{eq:cartesian_to_spherical_r}
\end{equation}

\begin{equation}
    \theta = \text{arccos}\left(\frac{z}{r}\right) \text{,}
    \label{eq:cartesian_to_spherical_theta}
\end{equation}

\begin{equation}
    \phi = \text{sgn}(y)\text{arccos}\left(\frac{x}{\sqrt{x^2+y^2}}\right) \text{,}
    \label{eq:cartesian_to_spherical_phi}
\end{equation}
\\
where $x$, $y$, and $z$ are coordinates of the normal $\vec{n}$, $r$ is the length of the normal $\vec{n}$, $\theta$ is the elevation angle, $\phi$ is the azimuth angle, and $\text{sgn}(\cdot)$ is the signum function.

\subsection{Assumptions}

Our work utilizes ray tracing method, which mimics real-world scenarios where a tranmission source emits numerous rays that interact with surrounding objects, bouncing off them indefinitely until losing intensity. 
% Light rays share numerous characteristics with wireless signals, differing primarily in frequency. Therefore, it is reasonable to employ ray tracing for modeling high-frequency signals, especially those operating at 28GHz, which is the focus frequency of our study.
% To optimize the utilization of our NVIDIA A100 GPU, we incorporate the maximum quantities of signal samples and reflections in our simulations. Notably, the transmitting power remains constant throughout the simulation, predetermined at the beginning of the simulation. Thus, the total power remains unaffected by the number of signal samples, with each sample carrying only a fraction of the total power.

We make an assumption that upon encountering an object, a ray undergoes specular reflection back into the environment (Figure \ref{figure:specular_reflection}). Each successive reflection diminishes the signal power, a reduction governed by a reflection coefficient. This coefficient is determined by both the ray's direction and the material composition of the surface it impinges upon \cite{rec_p2040}. Furthermore, the ray tracing method incorporates diffraction phenomenon \cite{keller1962geometrical, kouyoumjian1974uniform} to improve the simulation accuracy. 
% Nonetheless, we limit the number of diffractions encountered by individual rays due to the computational overhead associated with diffraction \cite{keller1962geometrical, kouyoumjian1974uniform}.

Several other assumptions are made. Firstly, all device positions are predetermined at the start of the simulation, a condition that can be realized in practical scenarios using tracking technologies such as Ultra Wideband (UWB) \cite{porcino2003ultra}. Additionally, we assume precise control over all tiles of a reflector device without addressing practical limitations. This enables us to freely adjust reflector tile rotation angles in the simulation to enhance UE path gains.

\subsection{Simulation}

% Intro
We investigate the downlink scenario, where an AP transmits information to a UE at a frequency of 28GHz. In this study, we leverage an open-source simulation wireless software library, Sionna by NVIDIA \cite{sionna}, to maximize received power at the UE location in 3D models. The software offers ray tracing simulations to generate coverage maps within a given geometry. The maximization of UE path gain, which is directly related to received power, is achieved by placing LFR devices at suitable locations. Based on a predefined configuration of AP, UE, and reflector devices, a 3D modeling software, particularly Blender \cite{blender}, can produce a desired 3D model. The geometry is then imported into the simulation software to generate coverage maps. 

In our simulation, we predefined the properties of each component of the 3D model. We configured a metallic LFR array consisting of 7x10 tiles (Figure \ref{figure:reflector}). Each metal tile measures 20 cm by 20 cm, resulting in a total device size of 1.4 m by 2.0 m. The walls, ceiling, and floor are made of concrete.

The relative permittivity and conductivity of all materials used in our simulation are specified according to guidelines provided by the International Telecommunications Union (ITU). Based on measurement data, the ITU derives frequency-dependent values for the real part of the relative permittivity, $\eta'$, and conductivity, $\sigma$, for many groups of materials\cite{rec_p2040}. Properties of materials that we use in our simulation are calculated as follows.

\begin{equation}
    \eta' = af^b,
    \label{eq:permittivity}
\end{equation}

\begin{equation}
    \sigma = cf^d,
    \label{eq:conductivity}
\end{equation}
\\
where $\eta'$ is dimensionless, $\sigma$ is in S/m, and $f$ is frequency in GHz. The values of $a, b, c$ and $d$ are presented in \ref{table:material_properties}.

\begin{table}[h]
\caption{Material Properties}
{\renewcommand{\arraystretch}{1.2}
\begin{center}
\begin{tabular}{ |c|c|c| } 
    \hline
    \textbf{Material} & Metal & Concrete \\
    \hline
    \textbf{a} & 1 & 5.24 \\ 
    \textbf{b} & 0 & 0 \\
    \hline
    \textbf{c} & $10^7$ & 0.0462 \\ 
    \textbf{d} & 0 & 0.7822 \\ 
    \hline
    \textbf{Frequency Range (GHz)} & 1-100 & 1-100  \\ 
    \hline
\end{tabular}
\end{center}
}
\label{table:material_properties}
\end{table}

% The workflow of our simulation is depicted in Figure \ref{figure:workflow}.

% \begin{figure}[!tbp]
%   \centering
%   \includegraphics[width=1\linewidth]{images/workflow-01.png}
%   \caption{Workflow of simulation}
%   \label{figure:workflow}
% \end{figure}

% Blender

\begin{figure}[h]
  \begin{subfigure}[b]{0.192\textwidth}
    \includegraphics[width=\textwidth]{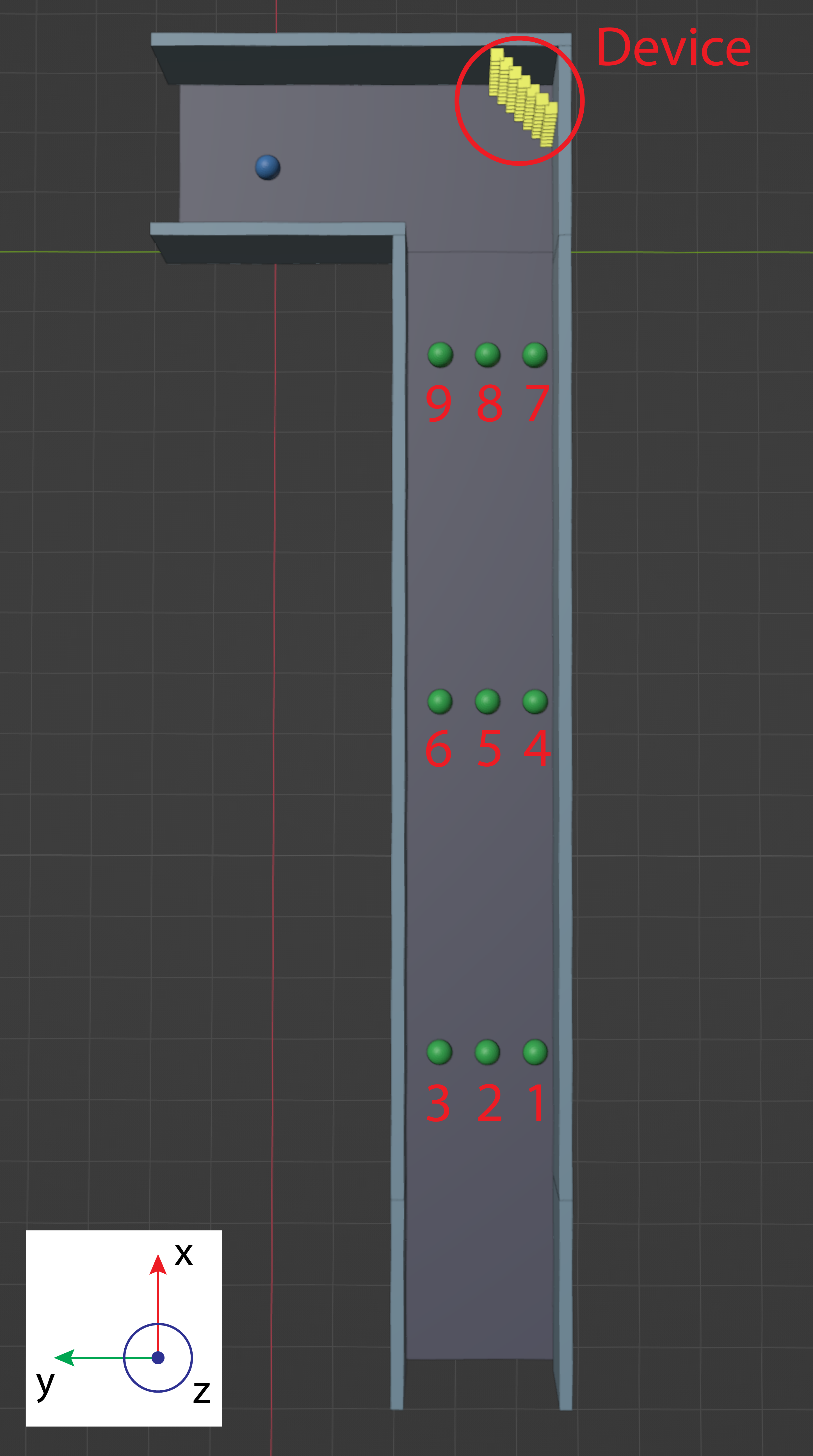}
    \caption{L-shape}
    \label{figure:hallway_l}
  \end{subfigure}
  \hfill
  \begin{subfigure}[b]{0.275\textwidth}
    \includegraphics[width=\textwidth]{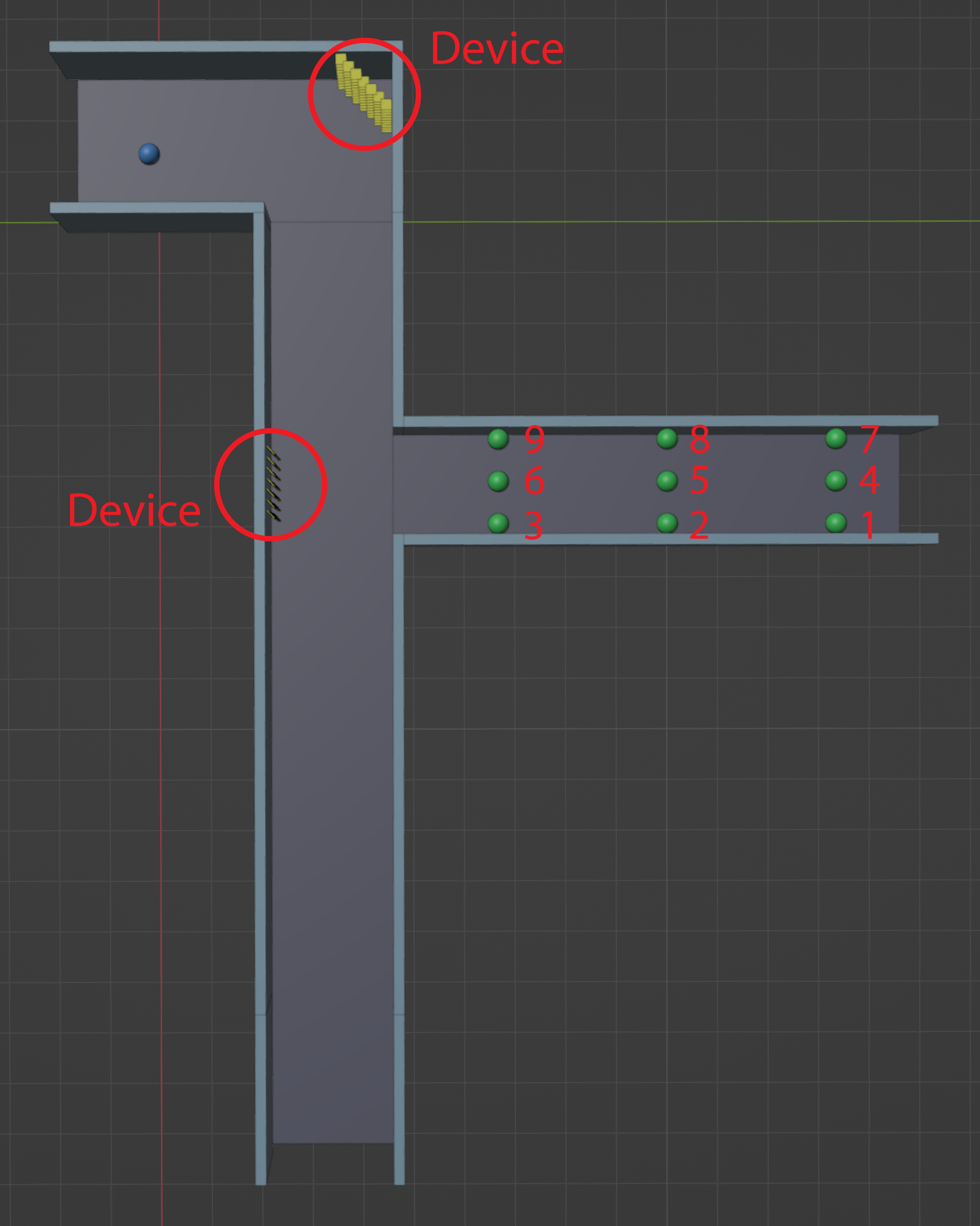}
    \caption{T-shape}
    \label{figure:hallway_t}
  \end{subfigure}
  \caption{Two hallway setups featuring AP and UE positions. The AP is represented by the blue sphere, while green spheres indicate the relocation of UE positions. For clarity, UE locations are distinguished by numbers.}
  \label{figure:hallway}
\end{figure}

We examine two primary cases: an L-shaped hallway and a T-shaped hallway (see Figure \ref{figure:hallway}). In both cases, the location of the AP remains fixed at the same location. We adjust the position of the UE to move through sections of the hallway where direct line-of-sight (LOS) paths are absent. The AP position is depicted in blue, whereas the locations of UEs are illustrated in green. 

The L-shaped case (Figure \ref{figure:hallway_l}) is designed as a fundamental case study to evaluate the efficacy of a single mechanical reflector array. Within the L-shaped case, we analyze three sub-cases: (1) a scenario without any reflector, (2) a scenario with a simple reflector, and (3) a scenario focusing on the beam-focusing technique. The first sub-case, without any reflector devices, establishes a baseline, representing a typical environment where no supporting devices are utilized to improve the received power at the UE. In contrast, the second and third sub-cases involve the placement of reflector devices at the hallway corner. In the simple reflector sub-case, all tiles within the reflector array rotate solely around the z-axis, each with an identical rotation angle of $45^o$. Conversely, the beam-focusing sub-case concentrates on creating multiple reflected paths through the reflector array, namely the AP-Reflector-UE paths. Through adjustments in the rotation angle of its tiles using the technique discussed in Section \ref{subsection:beamfocusing}, the reflector array can establish numerous reflected paths to the UE, thereby guiding various beams toward a specified location.

We expand our simulation to include a more complex layout, namely the T-shaped hallway, in which we incorporate two reflector arrays, as illustrated in Figure \ref{figure:hallway_t}. Similar to the analysis conducted for the L-shaped hallway, we explore the coverage maps for three different sub-cases: (1) without any reflector, (2) with two simple reflector arrays, and (3) emphasizing our beam-focusing technique. The configurations for all scenarios mirror those of the L-shaped hallway. In both the second and third scenarios, two reflector arrays are employed, one positioned at the hallway corner and the other facing the extended corridor. In the second sub-case, which involves two simple reflectors, the reflector tiles are rotated by $45^\text{o}$ around the z-axis. On the other hand, the objective of the third sub-case is similar to that of the L-shaped layout, striving to establish multiple reflected paths between the AP and UE through the utilization of two reflector arrays. As the UE relocates, we adjust the configuration of the two arrays to ensure that each tile of the two arrays establishes at least one reflected path from the AP to the UE. This involves conducting two consecutive bisector calculations, detailed in Section \ref{subsection:beamfocusing} while accounting for specular reflection.

% both reflectors can adapt the rotation angles of their tiles, using the beamfocusing technique in section \ref{subsection:beamfocusing}, to optimize the received power at the UE.

% We configure the second device so that each tile of the second device forms at least one indirect path from the AP and the first reflector to the UE. For example, when the AP shoots a ray to a tile of the first device, the reflected ray will impinge upon a corresponding tile of the second reflector before being reflected to the UE. Each tile of the first device correspond to a tile of the second device. We perform two consecutive bisector calculation for this scenario under specular reflection to maximize the received power at the UE.

We generate various scenarios and geometrical adjustment of all objects. Guided by predefined parameters of the scenario such as the locations of AP and UE, along with desired techniques and scene configurations, we proceed with the adjustment of reflector arrays. Primarily, this involves calculating the rotation angle of tiles comprising reflector arrays. These adjustments are made prior to exporting the scene for coverage map simulation.

% Sionna
Following the setup with Blender, each scenario is subsequently simulated using Sionna, which offers ray tracing simulation for wireless communication. 
Internally, Sionna uses Monte Carlo approximation to calculate the received signal power of an area. It creates a 2D grid at a height of 1.5m above the ground. This grid contains many small units, called cells, each of which is used to accumulate all received power values at the cell's center point. For example, when a ray hits a cell within a grid, the received power is calculated and accumulated in the cell. The process of calculation and accumulation happens until the maximum bounce is reached or the ray does not hit any objects in the scene. In the end, all received power values in a cell are summed to produce a single value at the center of that cell.

% Users have various options to customize settings based on the hardware accelerators available. We configured the number of samples (rays) and reflections to be 10 million and 15, respectively, fully utilizing the computational capacity of a single NVIDIA A100.

\section{Results and Discussion}
\label{sec:results_and_discussion}

We utilize wireless simulations carried out with Sionna to generate coverage maps illustrating path gains across various scenarios. Our results indicate that employing the beam-focusing technique significantly improves path gain, thereby enhancing the RSS at the UE. Moreover, we demonstrate that the RSS in scenarios employing the beam-focusing technique exceeds that of alternative scenarios at various locations.

\subsection{Coverage Map}

% Show coverage maps for L-shape and T-shape scenarios.
% Show 3 different settings: no reflector, simple reflector (all tiles have the same rotation angles), configurable reflector with small tiles
% Show how indiviadually controll tiles maximize the received power at the UE location.

% L-shape hallway
\begin{figure}[h]
  \begin{subfigure}[b]{0.14\textwidth}
    \includegraphics[width=\textwidth]{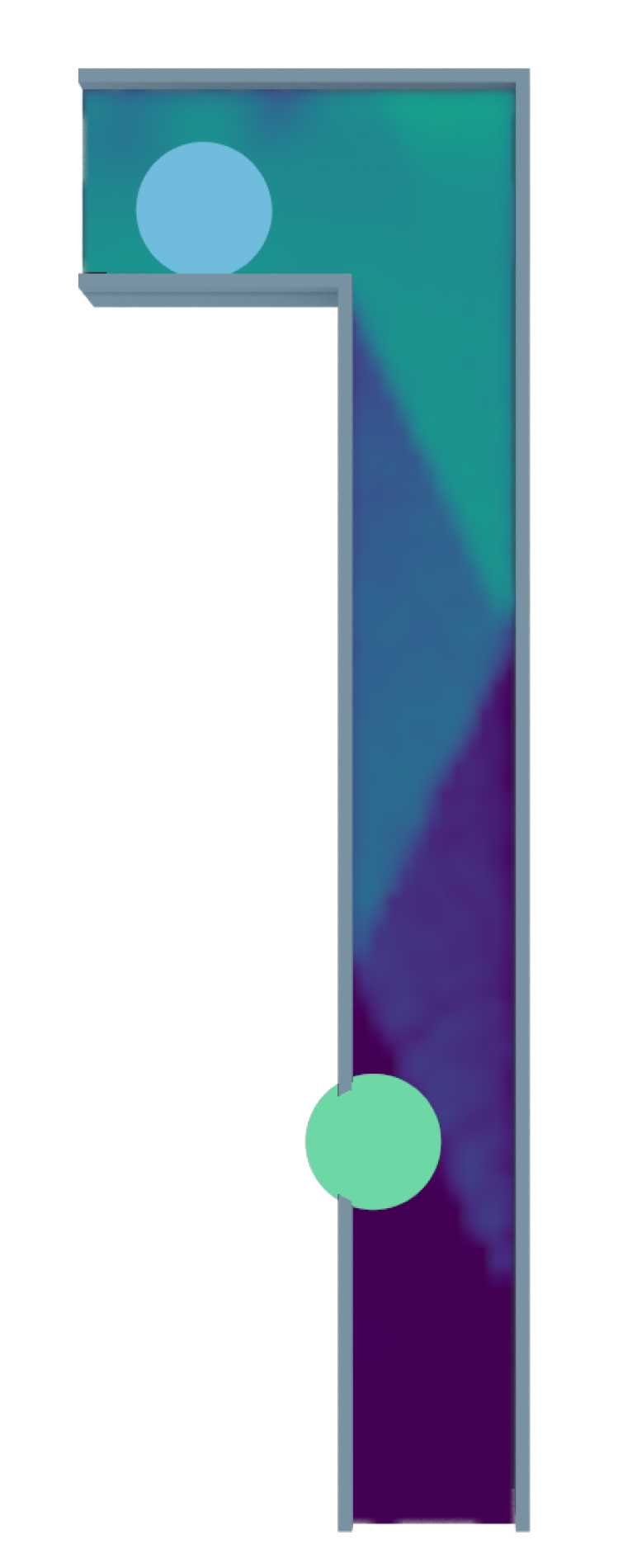}
    \caption{No reflector}
    \label{figure:hallway_l_no_reflector}
  \end{subfigure}
  \hfill
  \begin{subfigure}[b]{0.14\textwidth}
    \includegraphics[width=\textwidth]{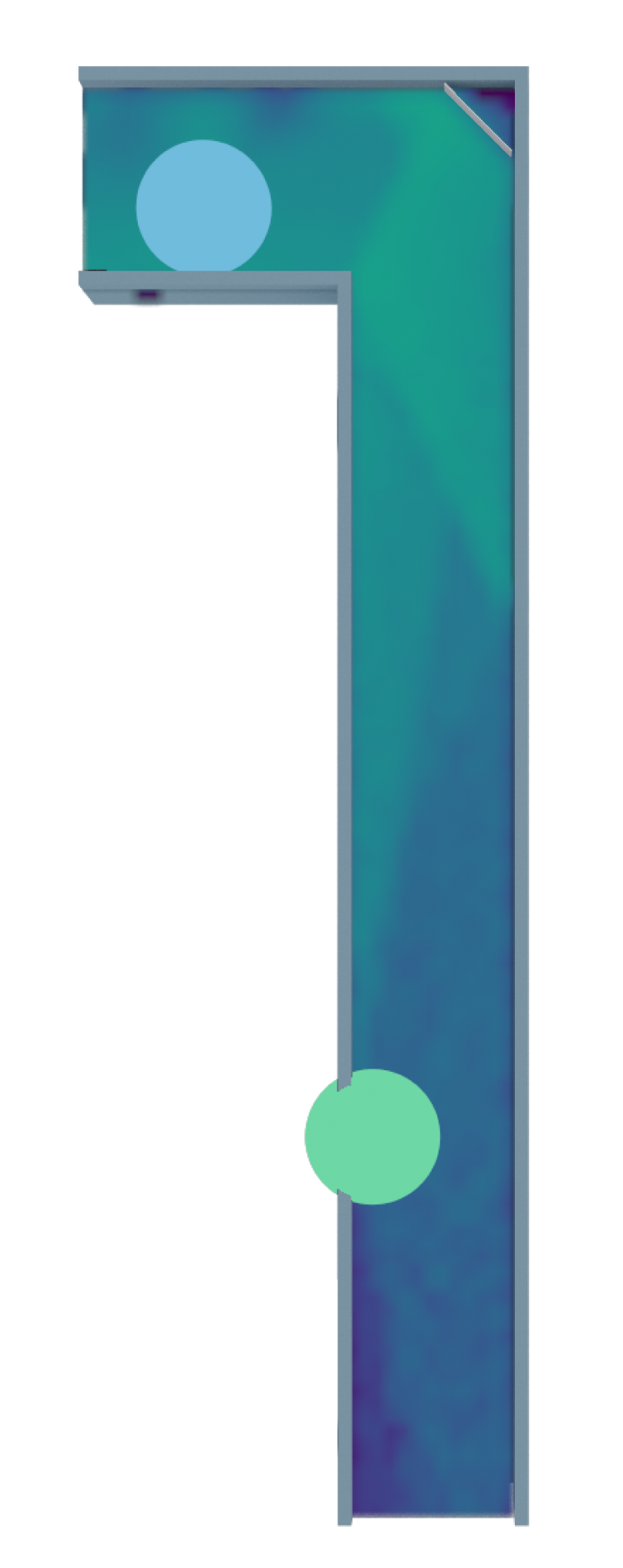}
    \caption{Simple reflector}
    \label{figure:hallway_l_simple_reflector}
  \end{subfigure}
  \hfill
  \begin{subfigure}[b]{0.185\textwidth}
    \includegraphics[width=\textwidth]{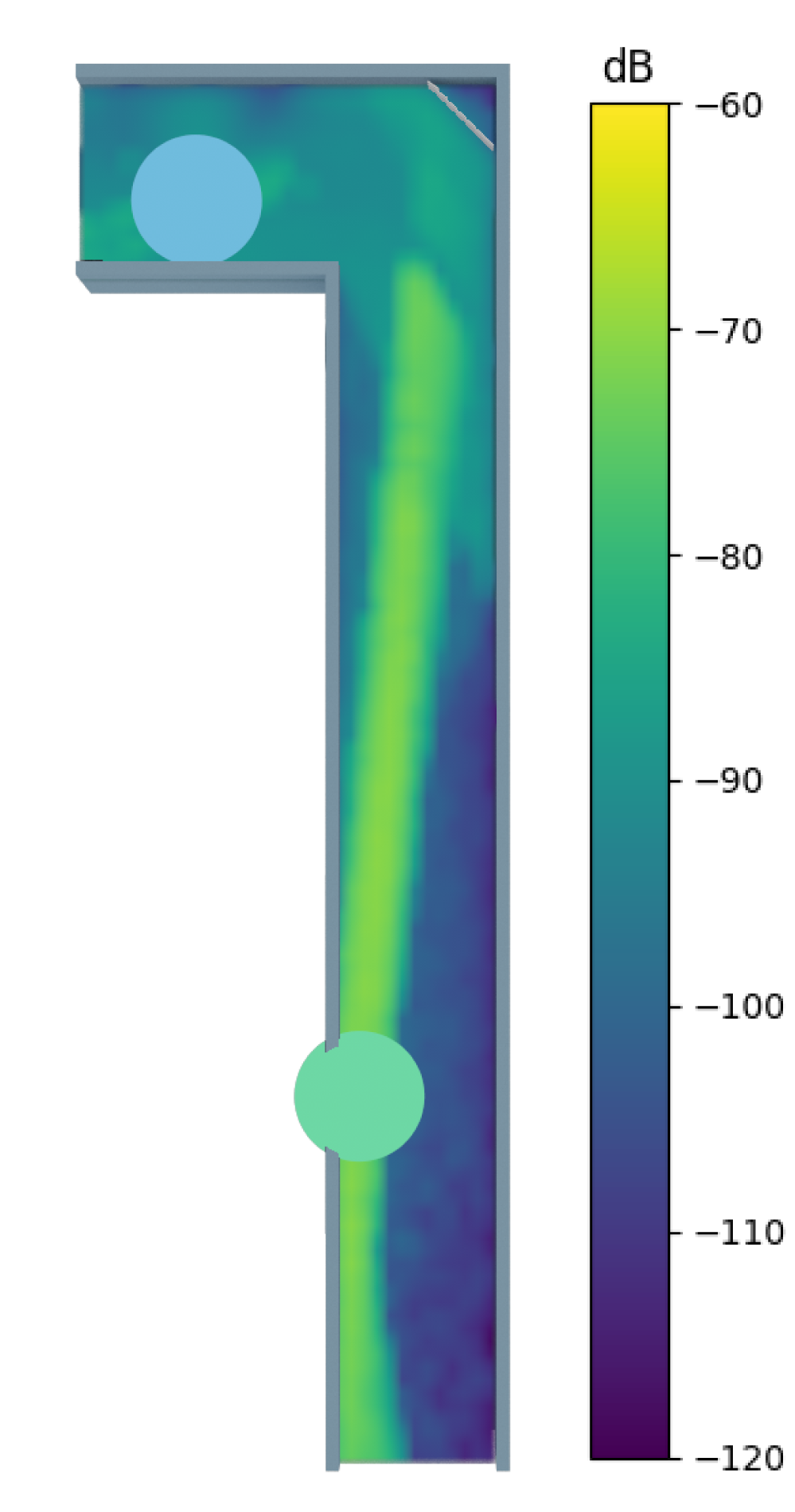}
    \caption{Beamfocusing}
    \label{figure:hallway_l_beamfocusing}
  \end{subfigure}
  \caption{Coverage maps of L-shape hallways displaying path gain represented in color.}
  \label{figure:hallway_l_cmap}
\end{figure}

% The coverage map is a 2D plane located 1.5m above the ground, while the scene is 3D. signal samples, reflected from the lower tiles of the device and upper part of the device, concentrate to the UE location at the height of 1.5m. That is why the path gain increases subtaintially for the beamfocusing sub-case.

% Therefore, when all beam focus to a single point, 

Coverage maps display path gain through a color map. As illustrated in Figure \ref{figure:hallway_l_cmap}, our simulation results indicate the efficacy of our reflector array. Figure \ref{figure:hallway_l_no_reflector} demonstrates that the signal strength remains unchanged regardless of the UE's position since there are no devices or objects to guide the signal transmission. Given the UE position in Figure \ref{figure:hallway_l_no_reflector}, the path gain at the UE is measured at -125.61 dB. On the other hand, Figure \ref{figure:hallway_l_simple_reflector} illustrates a marginal improvement of path gain at the UE to -99.54 dB, when employing a simple reflector. 

Figure \ref{figure:hallway_l_beamfocusing} demonstrates a significant enhancement in the path gain at the UE through the utilization of individual control for each tile on the reflector. With this approach, we improve the UE's path gain to -72.57 dB. This path gain, achieved with our technique, surpasses that of both the scene without any devices and the scene featuring a simple reflector. 
It is crucial to clarify the alteration in color observed in this specific scenario. The depiction of the coverage map represents a 2D x-y plane positioned at an elevation of 1.5 meters above ground level, while the surrounding environment exhibits 3D characteristics. Signals reflect off both the lower and the upper section of the array, ultimately converging at the UE location, precisely at a height of 1.5 meters. This convergence phenomenon leads to a significant amplification in path gain (resulting in color change) along the trajectories of all focused beams.

% T-shape hallway
% \begin{figure}[!tbp]
%   \begin{subfigure}[b]{0.145\textwidth}
%     \includegraphics[width=\textwidth]{images/no_reflector_T_hallway_00159-01.png}
%     \caption{No reflector}
%     \label{figure:hallway_t_no_reflector}
%   \end{subfigure}
%   \hfill
%   \begin{subfigure}[b]{0.145\textwidth}
%     \includegraphics[width=\textwidth]{images/simple_reflector_T_hallway_00159-01.png}
%     \caption{Simple reflector}
%     \label{figure:hallway_t_simple_reflector}
%   \end{subfigure}
%   \hfill
%   \begin{subfigure}[b]{0.175\textwidth}
%     \includegraphics[width=\textwidth]{images/beamfocusing_T_hallway_00159_color_map-01.png}
%     \caption{Beamfocusing}
%     \label{figure:hallway_t_beamfocusing}
%   \end{subfigure}
%   \caption{Coverage maps of T-shape hallways displaying path gain represented in color. Color scale is the same as Figure \ref{figure:hallway_l_cmap}.}
%   \label{figure:hallway_t_cmap}
% \end{figure}

\begin{figure}[h]
  \begin{subfigure}[b]{0.24\textwidth}
    \includegraphics[width=\textwidth]{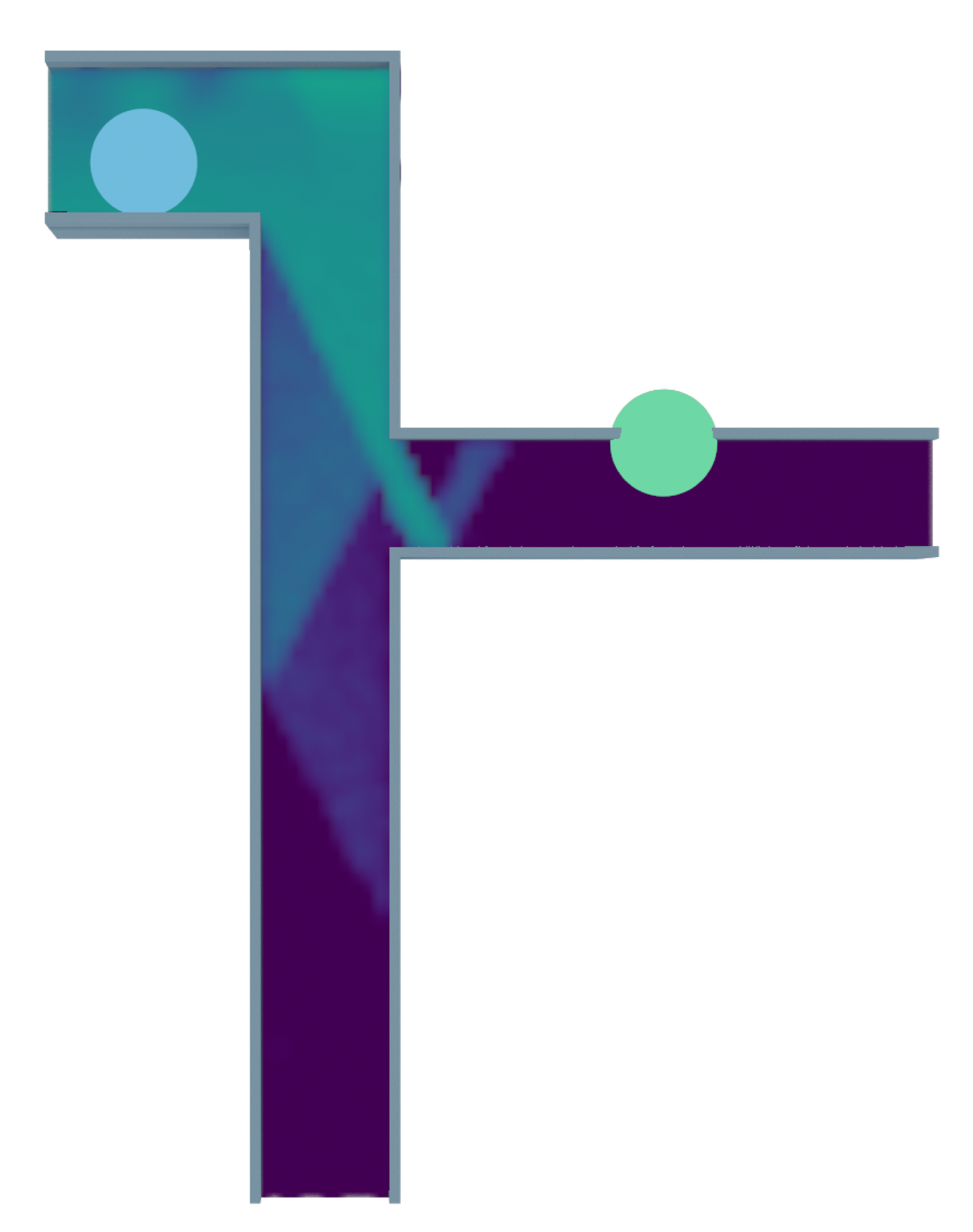}
    \caption{No reflector}
    \label{figure:hallway_t_no_reflector}
  \end{subfigure}
  \hfill
  \begin{subfigure}[b]{0.24\textwidth}
    \includegraphics[width=\textwidth]{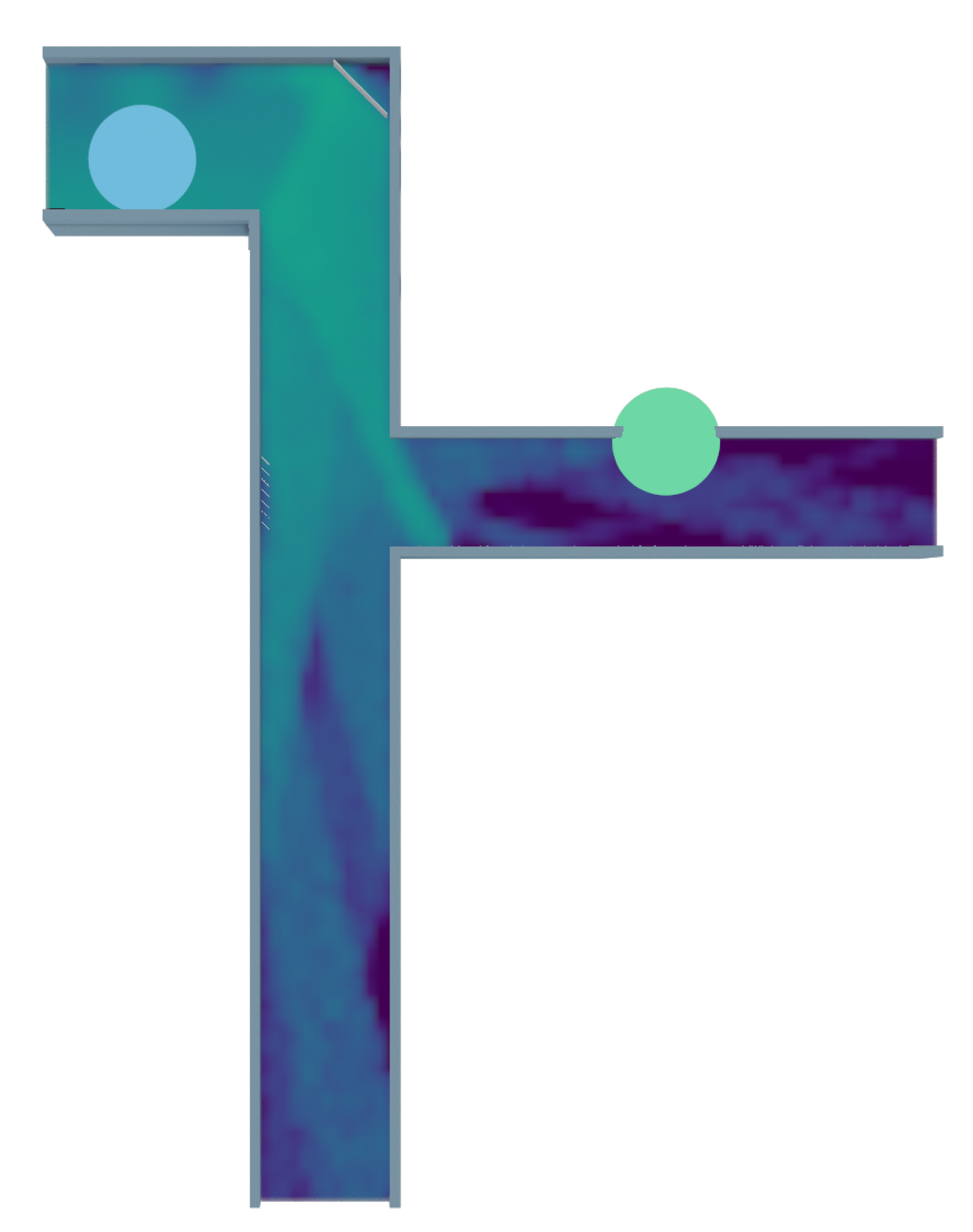}
    \caption{Simple reflector}
    \label{figure:hallway_t_simple_reflector}
  \end{subfigure}
  \hfill
  \begin{subfigure}[b]{0.30\textwidth}
    \includegraphics[width=\textwidth]{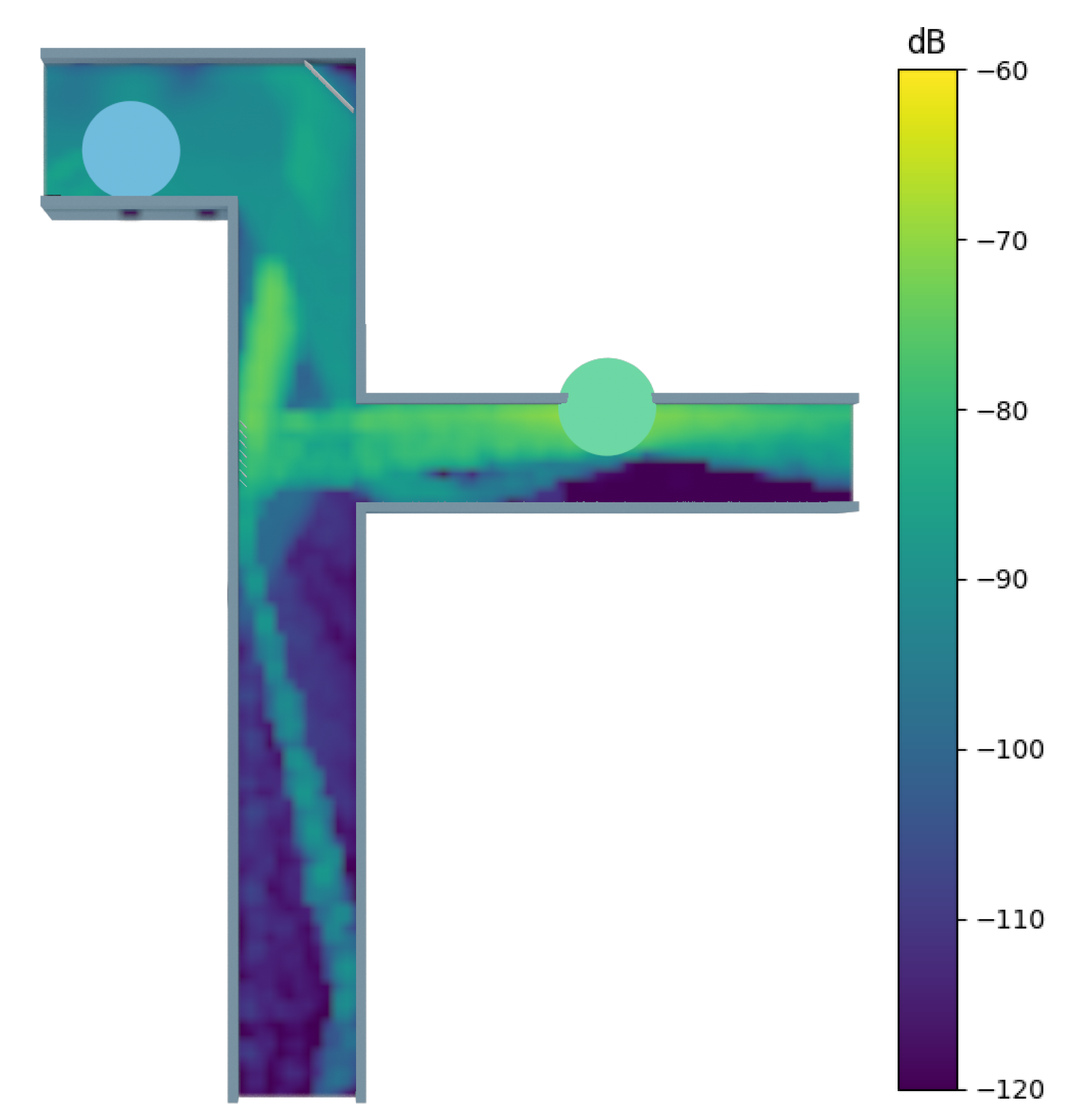}
    \caption{Beamfocusing}
    \label{figure:hallway_t_beamfocusing}
  \end{subfigure}
  \caption{Coverage maps of T-shape hallways displaying path gain represented in color}
  \label{figure:hallway_t_cmap}
\end{figure}

Similar to the findings from the L-shape hallway scenario, 
our simulation outcomes for the T-shaped hallways confirm the effectiveness of our reflector device, as depicted in Figure \ref{figure:hallway_t_cmap}. Figure \ref{figure:hallway_t_no_reflector} demonstrates that without any supporting devices, the path gain at the UE is estimated to be -156.35 dB. 
Meanwhile, Figure \ref{figure:hallway_t_simple_reflector} demonstrates an improvement in path gain at the UE with the integration of two simple reflectors, resulting in an increase to -109.20 dB. 
On the other hand, Figure \ref{figure:hallway_t_beamfocusing} depicts a substantial increase in path gain at the UE when complex reflectors are used. With the individual control of each tile to guide signal beams, we improve the UE's path gain to -68.88 dB.

As evident from Figure \ref{figure:hallway_l_cmap} and Figure \ref{figure:hallway_t_cmap}, the individual control of each tile in a reflector demonstrates the effectiveness in enhancing the path gain at the UE's location. The results, showing the proficiency of our method to track the UE's position while simultaneously improving its path gain, is presented at \cite{le_2024}.
% However, concentrating most of the signal towards the UE may result in diminished power at other locations within the hallway. This potential issue may arise particularly when employing multiple UE devices. However, we intend to address this challenge in our future work, particularly in scenarios involving multiple UEs.

\subsection{Received Signal Strength}
\begin{figure}[h]
  \begin{subfigure}[b]{0.24\textwidth}
    \includegraphics[width=\textwidth]{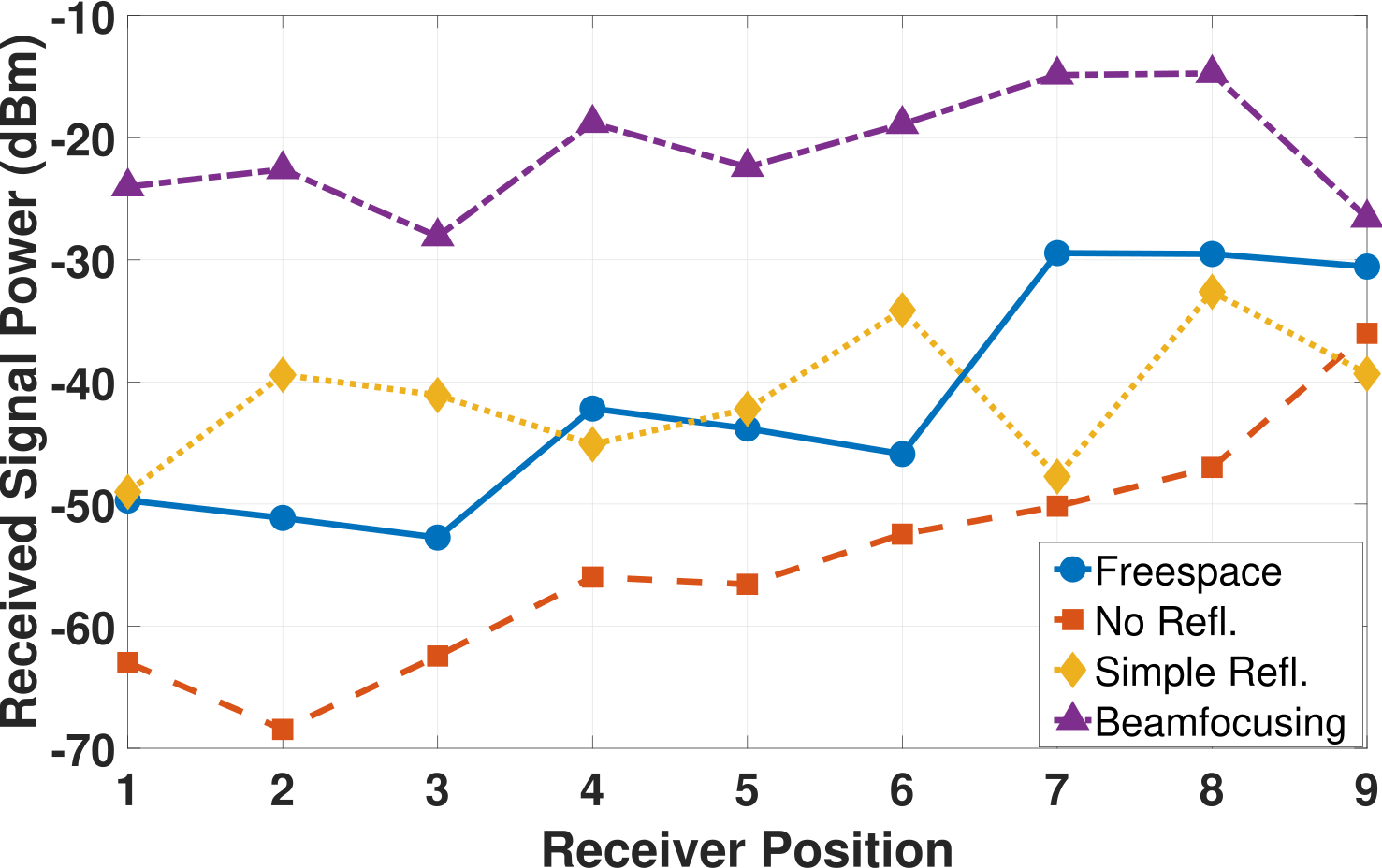}
    \caption{RSS for L-hallway}
    \label{fig:rss_L}
  \end{subfigure}
  \hfill
  \begin{subfigure}[b]{0.24\textwidth}
    \includegraphics[width=\textwidth]{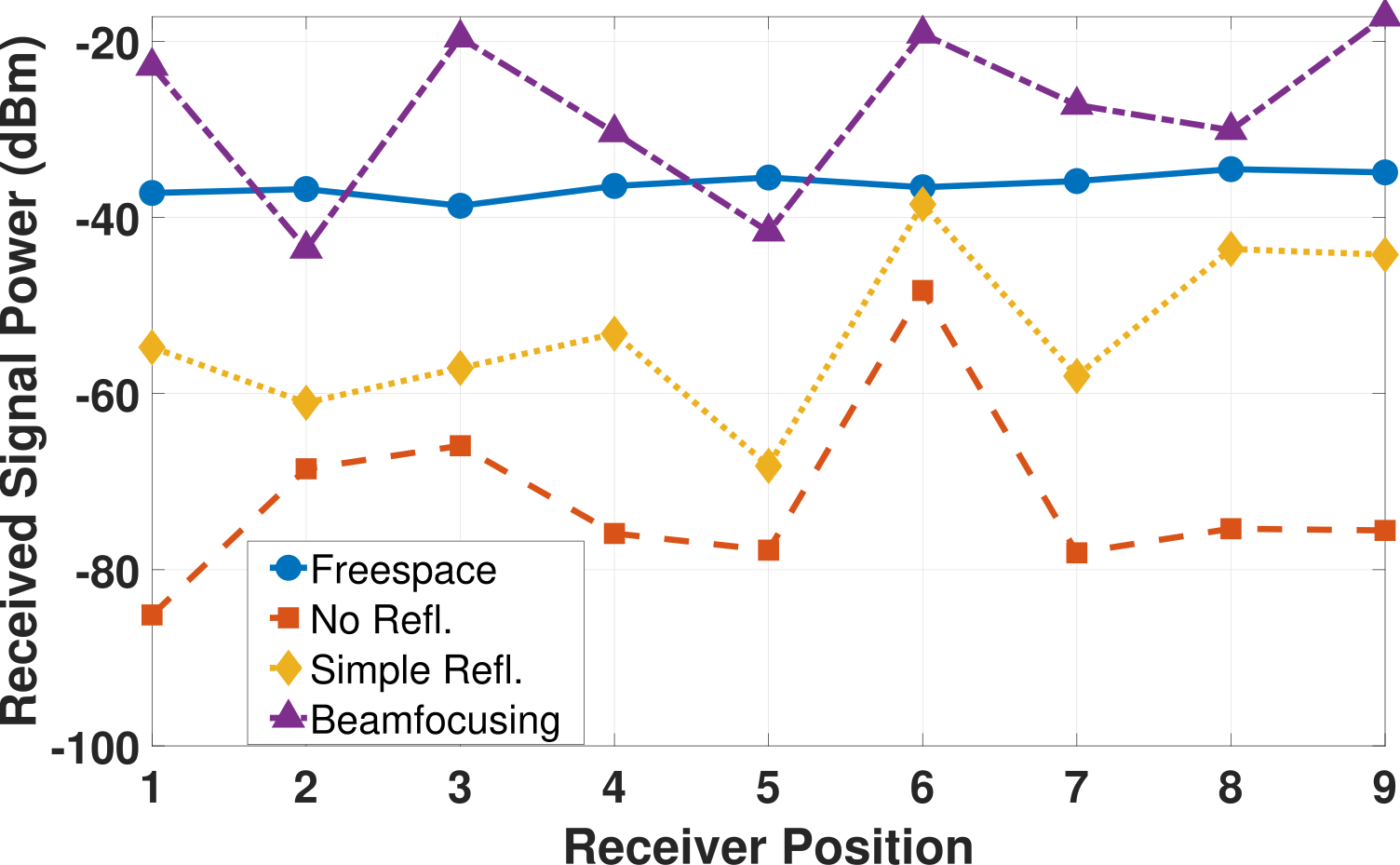}
    \caption{RSS for T-hallway}
    \label{fig:rss_T}
  \end{subfigure}
  \caption{Received signal strength (RSS) for nine different locations for each hallway}
  \label{fig:rss}
\end{figure}

Figure \ref{fig:rss} presents a comparative analysis of RSS, measured in dBm, when employing a transmitter with 10 watts transmit power across four distinct scenarios at nine predetermined locations within the area of interest, as depicted in Figure \ref{figure:hallway}. Notably, in Figure \ref{fig:rss_L}, the employment of a beam-focusing reflector array significantly enhances RSS across all locations, surpassing even free space transmission outcomes. Figure \ref{fig:rss_T} displays similar findings, with two exceptions at locations 2 and 5, where RSS with beam-focusing reflector arrays is lower than RSS under free space conditions. This discrepancy is currently attributed to certain multipath components destructively interfering with each other at these specific locations, resulting in reduced RSS with the beam-focusing reflector. Further investigation into this phenomenon is planned for future work. Lastly, it is noteworthy that even a simple specular reflective surface can have a non-negligible impact on RSS.

\subsection{Path to Practicality}

While the results of our proposed method are anticipated even without extensive research, this study employs simulations to quantify the path gain improvements achieved by using reflectors for beam focusing. The simulations show a path gain increase of at least 50 dB in both scenarios, indicating that this approach is promising for enhancing received signal strengths at the UE. Additionally, it has the potential to reduce power consumption in indoor environments, particularly for home users with stationary receiving devices, as it does not consume power once the reflector device is configured. Therefore, comparing our proposed method with conventional RIS implementations is not appropriate due to the latter's higher power consumption requirements.

To implement our reflector, we are investigating the use of MEMS mirrors as the primary tiles for the device. Each MEMS mirror tile is controlled by simple motors that allow limited movement in both horizontal and vertical directions. Specifically, each tile can adjust its orientation within a 20-degree range in both axes. The reflective surface of the MEMS mirrors is made of aluminum, which is cost-effective to manufacture and provides sufficient reflective properties for efficient beam-focusing. In addition, the quick response times and precise control capabilities of the MEMS mirrors make them ideal for dynamic beam steering, further enhancing the performance and adaptability of our reflector system. This approach aims to optimize the device's functionality while maintaining low production costs and ensuring easy integration into existing indoor environments.

\section{Conclusion and Future Work}
\label{sec:conclusion}

In this proof-of-concept study, we showcase the effectiveness of our Metallic Linear Fresnel Reflector Array in significantly improving signal quality by maximizing path gain at the UE. Utilizing geometric principles, we adjust the tiles of the reflector array to guide signal beams toward a desired destination. Results from simulation conducted in both L-shaped and T-shaped hallway scenarios validate the efficacy of our approach. Implementation of our reflector arrays leads to a substantial increase in path gains in both scenarios. Furthermore, we demonstrate that employing beam-focusing techniques can further improve path gain at the UE, often surpassing free-space path loss in the majority of cases.

However, this research is still in its initial stage, and there are various opportunities for improvement. For instance, practical systems entail physical constraints on the degree to which a tile can rotate. Given the complexity of the constraint control problem, we propose leveraging deep reinforcement learning to autonomously adjust reflector tiles, aiming to optimize received power at the UE location while respecting imposed constraints. Furthermore, physical experiments will be conducted to validate our simulation findings.

\section*{Acknowledgment}

This material is based upon work supported in part by the U.S. Department of Energy, Office of Science, Office of Advanced Scientific Computing Research, Early Career Research Program under Award Number DE-SC-0023957, and in part by the National Science Foundation under Grant No. 2323300

\bibliography{main}
\bibliographystyle{IEEEtran}

\end{document}